\documentclass[runningheads,a4paper]{llncs}

\usepackage{tablefootnote}
\usepackage{amssymb}
\usepackage{graphicx}
\usepackage{url}
\usepackage{amsmath}
\usepackage{listings}
\usepackage[linesnumbered,ruled,vlined]{algorithm2e} 
\usepackage[utf8]{inputenc} 
\usepackage[OT2,T1]{fontenc} 

\newcommand\textcyr[1]{{\fontencoding{OT2}\fontfamily{wncyr}\selectfont #1}}

\newenvironment{itemize2}
       {\begin{itemize}
                \vspace{-0.25em}
                 \setlength{\abovedisplayskip}{0pt}
                 \setlength{\belowdisplayskip}{0pt}
                 \setlength{\itemsep}{4pt}
                 \setlength{\parskip}{0pt}
                 \setlength{\parsep}{0pt}
                 \setlength{\topsep}{0pt}
                 \setlength{\partopsep}{0pt}
         }
         {\vspace{-0.25em}
         \end{itemize}}

\newenvironment{enumerate2}
       {\begin{enumerate}
                \vspace{-0.25em}
                 \setlength{\abovedisplayskip}{0pt}
                 \setlength{\belowdisplayskip}{0pt}
                 \setlength{\itemsep}{4pt}
                 \setlength{\parskip}{0pt}
                 \setlength{\parsep}{0pt}
                 \setlength{\topsep}{0pt}
                 \setlength{\partopsep}{0pt}
         }
         {\vspace{-0.25em}
         \end{enumerate}}

\urldef{\mailsa}\path|{a.panchenko}@digsolab.com|    
\newcommand{\keywords}[1]{\par\addvspace\baselineskip
\noindent\keywordname\enspace\ignorespaces#1}

\begin{document}

\mainmatter  

\title{Large-Scale Parallel Matching of \\ Social Network Profiles }


\titlerunning{Large-Scale Parallel Matching of Social Network Profiles}

\author{Alexander Panchenko\inst{1} \and Dmitry Babaev\inst{2} \and Sergei Obiedkov\inst{3}  }

\authorrunning{A. Panchenko and D. Babaev and S. Objedkov}

\institute{
TU Darmstadt, FG Language Technology, Darmstadt, Germany  
\url{panchenko@lt.informatik.tu-darmstadt.de}
\and
Tinkoff Credit Systems Inc., Moscow, Russia  \\
\url{dmitri.babaev@gmail.com.}
\and 
National Research University Higher School of Economics, Moscow, Russia \\
\url{sergei.obj@gmail.com}
}

%
%

\toctitle{ \ldots } 
\tocauthor{ \ldots }
\maketitle

\begin{abstract}

A profile matching algorithm takes as input a user profile of one social network and returns, if existing, the profile of the same person in another social network. Such methods have immediate applications in Internet marketing, search, security, and a number of other domains, which is why this topic saw a recent surge in popularity. 

In this paper, we present a \textit{user identity resolution} approach that uses minimal supervision and achieves a precision of 0.98 at a recall of 0.54. Furthermore, the method is computationally efficient and easily parallelizable. We show that the method can be used to match \textit{Facebook}, the most popular social network globally, with \textit{VKontakte}, the most popular social network among Russian-speaking users. 

\end{abstract}


\section{Introduction}

Online social networks enjoy a tremendous success with general public. They have even become a synonym of the Internet for some users. While there are clear global leaders in terms of the number of users, such as Facebook\footnote{\url{http://www.facebook.com}}, Twitter\footnote{\url{http://www.twitter.com}} and LinkedIn\footnote{\url{http://www.linkedin.com}}, these big platforms are constantly challenged by a plethora of niche and/or local social services trying to find their place on the market. For instance, VKontakte\footnote{\url{http://www.vk.com}} is an online social network, similar to Facebook in many respects, that enjoys a huge popularity among Russian-speaking users.

 Current situation leads to the fact that many users are registered in several social networks. People use different services in parallel as they provide complimentary features and user bases. For instance, one common pattern for Russian-speaking users is to communicate with Russian-speaking peers with help of Vkontakte and with foreign friends with help of Facebook. Another common pattern is to use LinkedIn for professional and Facebook for private contacts.

Publicly available user information can help in building the next generation of personalised web services, such as search, recommendation systems, targeted marketing, and messaging, to name a few. For instance, Bartunov et al.~\cite{bartunov2012joint} suggest to use profile matching to perform automatic contact merging on mobile phones. Actually, a similar technology is already integrated in the Android mobile operative system\footnote{https://www.android.com}. On the other hand, profile information may be subject to de-anonymization attacks, undesirable for a user~\cite{balduzzi2010abusing,wondracek2010practical,goga2013large}. No wonder several researchers from information retrieval and security communities recently tried to study methods of user profile correlation across online social networks~\cite{sironi2012automatic,veldman2009matching,narayanan2009anonymizing,raad2010user,malhotra2012studying,goga2013large,jain2013seek}.

As information about a single user can be scattered across different networks, integration of data from various platforms can lead to a more complete user representation. Therefore, in many applications it makes sense to build an \textit{integral profile}, featuring information from several sources. In order to do so, it is necessary to perform \textit{user identity resolution}, i.e., to find the same person across various networks. In this paper, we propose a simple, yet efficient method for matching profiles of online social networks.

The contribution of our work is two-fold:
\begin{enumerate2}
\item We present a new method for matching profiles of social networks. The method has only four meta-parameters. Unlike most existing approaches (see Section~\ref{sec:related}), the method is easily parallelisable and can be used to process the profiles from an entire social network in a matter of hours. We provide an open-source implementation of the method.\footnote{\url{https://github.com/dmitrib/sn-profile-matching}}

\item We present results of the largest matching experiment to date known to us. While most prior experiments operated on datasets ranging from thousands to hundreds of thousands of profiles, we performed a match of 3 million profiles of Facebook (FB) to 90 million profiles of VKontakte (VK), demonstrating that third parties can perform matching on the scale of entire social network. To the best of our knowledge, we are the first to present a matching of FB to VK. 

\end{enumerate2}

\section{Related Work}
\label{sec:related}

\subsection{Profile Matching}

Bartunov et al.~\cite{bartunov2012joint} developed a probabilistic model that relies on  profile attributes and friendship links. The algorithm was tested on roughly 2 thousand Twitter users and 9 thousand Facebook users. The method achieves F-measure up to 0.89 (precision of 1.0 and recall of 0.8). However, the this is a \textit{local} identity resolution method,  that requires profiles to be ego-networks of the seed user. From the other hand, in this paper we present a \textit{global} identity resolution method that can potentially match any user of one network with any user of another network.  

Veldman~\cite{veldman2009matching} conducted a set of extensive experiments with profile matching algorithms. She used profile similarity metrics based on both attributes (name, email and birth date) and friendship relations. The author performed experiments on 2 thousand profiles of LinkedIn and Hyves social networks. 

Malhotra et al.~\cite{malhotra2012studying} used 30 thousand of paired Twitter and LinkedIn profiles to train several supervised models based on attributes, such as name, user id and location. The authors report an F-measure up to 0.98 with precision up to 0.99. 

Sironi~\cite{sironi2012automatic} also used supervised models based on features stemming from similarity of profile attributes. This experiment was done on 34 thousand of Facebook, Twitter and LinkedIn profiles where 2 thousand were paired. Their approach yields precision and recall around 0.90.

Narayanan and Shmatikov~\cite{narayanan2009anonymizing} proposed an approach that establishes connections between users based on their friendship relations. This incremental method requires a small initial number of matched profiles and access to a graph of friendship links. The authors used the method to match 224 thousand Twitter users with 3.3 million Flickr users and observed an error rate of 12\%. 

Balduzzi et al.~\cite{balduzzi2010abusing} showed that matching can be done effectively based on email addresses. 

Jain et al.~\cite{jain2013seek} developed a system that takes as input a Twitter account and finds a corresponding Facebook account. The system relies on profile, content, self-mention and network-based similarity metrics. 

Goga et al.~\cite{goga2013large} present a comprehensive study on profile matching technology. The authors try to correlate accounts of Facebook, Twitter, Google+, Flickr, and MySpace to check a feasibility of a de-anonymization  attack. They show that up to 80\% of Twitter, Facebook and Google+ profiles from their ground truth can be matched with a nearly zero false positive rate. Their matching method is based on features extracted from user names, locations and pictures unified with help of a binary classifier taking as input two profiles. Two key differences of this method from ours are the following. First, Goga et al.~\cite{goga2013large} perform no candidate selection. Therefore, in this approach all pairwise comparisons should be done, which is not efficient if one deals with the entire social network. Second, this approach uses no features based on friends similarity, which are core of our approach.

\subsection{Name Similarity Matching}

Our method heavily relies on name similarity matching. In its simplest form a name can be considered as a string. There is a large body of literature on how to define string similarity \cite{boytsov2011indexing} and use it to extract similar names from a data set with some works focusing specifically on personal names; see a survey and experimental comparison in \cite{du2005approximate}. According to this survey, one of the best algorithms for approximate name matching is the algorithm from \cite{navarro2003matchsimile}, which uses a prefix tree to efficiently compute the Levenshtein distance. In \cite{lisbach2013linguistic}, three generations of name matching methods are identified, with only third-generation methods showing good results in terms of both precision and recall. 

\section{Dataset}
Two social networks were used in our profile matching experiment. One is the biggest Russian social network VKontakte; the other is Facebook, which is also very popular among Russian-speaking users.

In our experiments, we used publicly available data from VK and FB. The matching algorithm is based on name similarity and the friendship relation: each profile is represented by the first and/or last name of the user and by a list of names of his or her friends in the social network. No other characterising features of profiles were used.

\subsection{VKontakte}

We collected about 90 million VK profiles that set Russia as their current location. We gathered first and second name of each user along with list of her friends using the ``users.get'' method of the social network API\footnote{ \url{https://vk.com/dev/users.get} }. Therefore, we can assume that in our experiment VK friend lists are \textit{complete}.

\subsection{Facebook}
We deal with 3 million public Facebook profiles from Russia. User's name can be obtained via the official API\footnote{\url{https://developers.facebook.com/docs/graph-api}}, but not list of her friends. That is why friend lists were generated from events displayed in user's feed. Users A and B were considered as friends if a message ``A and B are now friends'' appeared in feeds of A and B. Profile feeds were collected via the ``user/feed'' method of the FB API. The problems with this approach is that (1) access to users's feed can be restricted by privacy settings; (2) one need to download all wall posts to gather  list of friends, which is not always possible due to API restrictions and requires multiple API calls. Therefore, we should assume that in our experiment FB friend lists are \textit{incomplete}.

\subsection{Test data}
\label{sec:testdata}

VKontakte provides a field where a used can specify a link to her FB page. We gathered about 850 thousand known VK-FB profile pairs. However, only 92,488 Facebook users were found in our Facebook dataset out of these 850 thousand profiles. These pairs were used as a ground truth to check correctness of the matching algorithm. A subset of the test data used in our experiments is publicly available\footnote{\url{https://github.com/dmitrib/sn-profile-matching}}.

\subsection{Name romanisation}

Names of Russian FB and VK users can be spelled in both Latin and Cyrillic alphabets i.e. ``Alexander Ivanov'' or ``\textcyr{Александр Иванов}''. To enable correct name matching, all user names in both networks were converted to Latin script using the Russian-Latin BGN transliteration rules\footnote{\url{http://earth-info.nga.mil/gns/html/romanization.html}}.

\section{Profile matching algorithm}

The algorithm consists of three phases:
\begin{enumerate2}
\item \textit{Candidate generation}. For each VK profile we retrieve a set of FB profiles with similar first and second names. 
\item \textit{Candidate ranking}. The candidates are ranked according to similarity of their friends.
\item \textit{Selection of the best candidate}. The goal of the final step is to select the best match from the list of candidates.
\end{enumerate2}

Each profile from VK network is processed independently and hence this operation can be easily parallelised (we rely on MapReduce framework\footnote{\url{http://hadoop.apache.org}}). It is possible to perform matching in both directions (VK$\rightarrow$FB and FB$\rightarrow$VK). However, all profiles from the target network must be stored at each computational node. Therefore, direction of matching VK $\rightarrow$ FB minimises the memory footprint of such nodes. Below we describe each step of the method in detail.

\subsection{Candidate generation}

It is computationally inefficient to calculate similarity of each VK profile with each  FB profile. This operation would require about $1.3\cdot10^{20}$ pairwise comparisons. This first step reduces the search space retrieving FB users with names similar to the input VK profile. Two names are considered similar if the first letter is the same and the edit distance~\cite{lisbach2013linguistic} between names is less than two. This should be true for both first and last names. 

We use an index based on Levenshtein Automata~\cite{Schulz02faststring} to perform fuzzy match between a VK user name and all FB user names. In particular, we relied on the Lucene implementation of this approach\footnote{\url{org.apache.lucene.util.automaton.LevenshteinAutomata}}.

However, the edit distance does not provide a complete solution for name matching, since many first names have several rather different variants, e.g., ``Robert'' and ``Bob'', or ``Mikhail'' and ``Misha''. One way to address this problem is to use a dictionary of proper names prepared by linguists, e.g., \cite{petrovsky1966dictionary}, to decide whether two names are  synonyms. However, such dictionaries often skip some name variants. For example, the entry for the Russian name ``Alexander'' in \cite{petrovsky1966dictionary} includes ``Sanya'', but not ``Sanek''. In addition, they do not include variants based on similarity with names from other languages: e.g., ``Alejandro'' is not in the entry for ``Aleksandr'' in \cite{petrovsky1966dictionary}. 

Therefore, we decided to build our own dictionary using pairs of profiles known to belong to the same person. We do this by taking the transitive closure of the symmetric binary relation over names given by these pairs. Every two names from the same equivalence class are considered to be synonyms. The fundamental deficiency of this approach is that, being a variant is not an equivalence relation, since transitivity does not always hold. For example, two different Russian names, ``Alexander'' and ``Alexey'', are often abbreviated as ``Alex''. With our approach, this results in declaring ``Alexander'' and ``Alexey'' variants of each other, which they are not. Nevertheless, we let this happen and use shared friends to disambiguate between persons erroneously declared to have similar names. 

Another problem is that some people use totally unrelated first names (such as ``Andrey'' and ``Vladimir'' or even ``Max'' and ``Irina'') in different networks. We solve this problem by  removing ``strange'' pairs based on the number of times such a pair occurs in the list (unique or infrequent pairs can safely be removed). The final list of synonym clusters was quickly checked manually.


While candidate generation step greatly reduces search space, a person that indicated different name or a pseudonym in two social networks will not be recognised with our approach. From the other hand, in this situation a person is probably prefers to hide his or her identity and therefore it is more appropriate to perform no matching for this user at all.

\subsection{Candidate ranking}

The higher the number of friends with similar names in VK and FB profiles, the larger the similarity of these profiles. Two friends are considered to be similar if:

\begin{itemize2}
\item First two letters of their last names match, and
\item The similarity between their first names and the similarity between their last names are both greater than thresholds $\alpha$ and $\beta$, correspondingly. We empirically set $\alpha$ to 0.6 and $\beta$ to 0.8. String similarity $sim_{s}$ is calculated as follows:
$$ 
sim_{s}(s_i,  s_j) =  1-  \frac{lev(s_i, s_j)}{\max(|s_i|, |s_j|)},
$$
where $lev$ is edit distance of string $s_i$ and $s_j$. At this step we use the standard algorithm for calculation of Levenstein distance\footnote{org.apache.lucene.search.spell.LevensteinDistance}, not Levenstein Automata.

\end{itemize2}

Matching friends with rare names should be weighted higher than a match of friend with matching friends with common names. Indeed, two unrelated profiles can easily have several friends with similar common names.

Probability of a user with first name $s^f$ and second name $s^s$, provided than these events are independent is $P(s^f,s^s) = P(s^f)P(s^s) = \frac{|s^f|}{N}\frac{|s^s|}{N}$. Here $|s^f|$ and $|s^s|$ are frequencies of respectively first and second names and $N$ is the total number of profiles. Thus, expectation of name frequency equals to $\frac{|s^f| \cdot |s^s|}{N}$. In our approach, contribution of each friend to similarity $sim_p$ of two profiles $p_{vk}$ and $p_{fb}$ is inverse of name expectation frequency, but not greater than one:  
$$
sim_p (p_{vk}, p_{fb}) = \sum_{ j : sim_s(s^f_i,s^f_j) > \alpha \wedge sim_s(s^s_i, s^s_j) > \beta }  \min (1, \frac{N}{ |s^f_j| \cdot |s^s_j|  }).
$$

Here $s^f_i$ and $s^s_i$ are first and second names of a VK profile, correspondingly, while $s^f_j$ and $s^s_j$ refer to a FB profile.

\subsection{Best candidate selection}

FB candidates are ranked according to their similarity $sim_p$ to an input profile $p_{vk}$. There are two thresholds the best candidate $p_{fb}$ should pass to match:

\begin{itemize2}

\item its score should be higher than the \textit{similarity threshold} $\gamma$:
$$
sim_p(p_{vk},p_{fb}) > \gamma.
$$ 

\item it should be either the only candidate or score ratio between it and the next best candidate $p'_{fb}$ should be higher than the \textit{ratio threshold} $\delta$:
$$
\frac{sim_p(p_{vk},p_{fb})}{sim_p(p_{vk},p'_{fb})} > \delta.
$$
\end{itemize2}


The $\delta$ threshold enforces the fact that a VK user has only one account in FB. On the other hand, one FB profile can be linked with several VK profiles. Still, in this case only the match with the highest score is kept.

\begin{figure}
\centering
\includegraphics[width=0.7\textwidth]{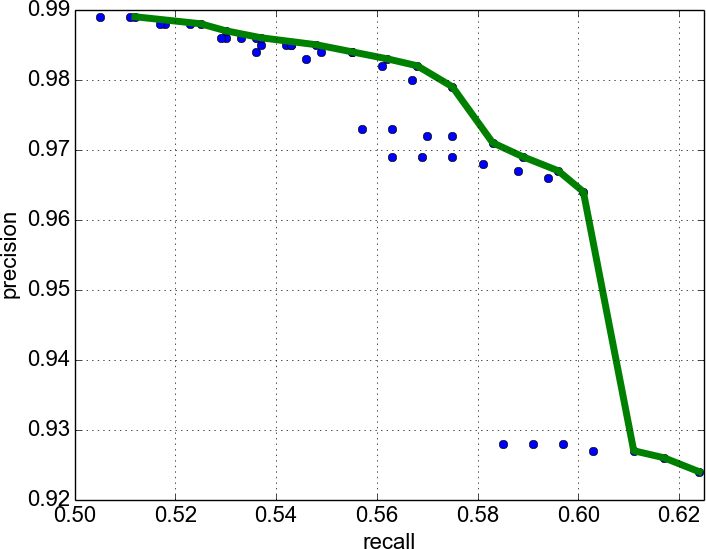}
\caption{Precision-recall plot of our matching method. Here we perform a grid search of two method parameters: profile \textit{similarity threshold} $\gamma \in [1;4]$ and profile similarity \textit{ratio threshold} $\delta \in [3;6]$. The bold line denotes the best precision at a given recall. }
\label{fig:pr}
\end{figure}

\section{Results and Discussion}

We performed matching of VKontakte and Facebook profiles (c.f. Table~\ref{tab:stat}) with the approach described above. Results of the candidate ranking step were saved. At this point, we conducted a series of experiments varying the similarity threshold $\gamma$ and the ratio threshold $\delta$. Results of these experiments in terms of precision and recall with respect to the test collection (see Section~\ref{sec:testdata} are presented in Figure~\ref{fig:pr}. The bold line denotes the best precision at certain level of recall.

\begin{table}
\centering

\caption{Statistics of VKontakte and Facebook. }

\begin{tabular}{|l|r|r|} \hline

 & \textbf{VKontakte} & \textbf{Facebook} \\ \hline \hline
Number of users in our dataset & 89,561,085 & 2,903,144 \\ 

Number  of Russian-speaking users\tablefootnote{\scriptsize \url{http://www.comscore.com/Insights/Data-Mine/Which-Sites-Capture-The-Most-Screen-Time-in-Russia} and \url{http://vk.com/about} provide statistics on number of Russian-speaking users.}  & 100,000,000 & 13,000,000 \\ 
User overlap & 29\% & 88\% \\ \hline
\end{tabular}
\label{tab:stat}
\end{table}

As one may observe, our method yields very good results achieving precision of 0.97 at recall of 0.58. Furthermore, a configuration of the approach yielding 99\% precision recalls roughly 50\% of relevant profiles. 

In order to perform the final matching of VK and FB we chose a version of the algorithm that provides precision of 0.98 and recall of 0.54 (see Table~\ref{tab:results}). Results were obtained in 4 hours on a Hadoop cluster with 100 nodes of type \texttt{m2.xlarge} (2 vCPU, 17 GB RAM) on the AWS EC2 cloud\footnote{\url{http://aws.amazon.com}}. The mentioned above configuration of the method mentioned above retrieved 644,334 VK profiles of FB users. Thus, we found corresponding VK pages of 22\% Facebook users present in our collection. 
 
\begin{table}

\centering
\caption{Matching of user profiles of Facebook and VKontakte social networks. The upper table presents four main parameters of our profile matching method. The lower part of the table presents results of the final matching of the two networks. }

\begin{tabular}{|l|l|} \hline
\textbf{Parameter} & \textbf{Value} \\ \hline \hline
First name similarity threshold, $\alpha$ & 0.8 \\ 
Second name similarity threshold, $\beta$ & 0.6 \\ 
Profile similarity threshold, $\gamma$ & 3 \\ 
Profile ratio threshold, $\delta$ & 5 \\ \hline \hline 
Number of matched profiles & 644,334 (22\% of 2,903,144 FB users) \\ 
Expected precision & 0.98 \\ 
Expected recall & 0.54 \\ \hline

\end{tabular}
\label{tab:results}
\end{table}

While our approach makes only few errors, reaching precision of 0.98, it is not able to match a significant fraction of 40-50\% of user profiles. The key factors hampering correct retrieval are the following: 

\begin{itemize2}
\item In our method, we perform fuzzy search with name synonyms that can lead to semantic drift. For instance, ``Maria'' is expanded with its alias ``Masha''. According to fuzzy search ``Masha'' and ``Misha'' are related. But the latter is a shortcut for ``Michael'' in Russian. 

\item Implementation of the Levenstein Automata used in our experiments retrieves candidates with distance lower or equal than two. Thus, people with long names and surnames can be missed during candidate generation.

\item People often intentionally indicate different names in two social networks or use different aliases. Our approach is not designed to identify and match such profiles.

\item  First letter mismatch. Different variants of the same name/surname in Russian can start from different letters.  Furthermore, transliteration can lead to such mismatches as well, e.g. surname ``\textcyr{Ефимов}'' can be spelled in Latin as ``Efimov'' or ``Yefimov''.  

\item People often use transliterated versions of their names in one network, but stick to the original Cyrillic versions in the other. The method always works with transliterated names, but our transliteration can be quite different from the one done by a user.

\item Due to nature of the friend collection method used, some FB friends can be absent in our dataset.   

\end{itemize2}

In order to improve performance of the method, one would need to tackle the problems mentioned above.  

\section{Conclusion}

In this paper, we presented a new user identity matching method. Unlike most previous approaches, our method is able to work on the scale of real online social networks, such as Facebook, matching tens of millions of users in several hours on a medium-sized computational cluster. The method yields excellent precision (up to 98\%). At the same time it is able to recall up to 54\% of correct matches. 

The method was used to perform the most largest-scale matching experiment up to date. We matched 90 millions of VKontakte users with 3 million of Facebook users. 

A prominent direction for the future work, is to use supervised learning  in order to improve \textit{candidate ranking}. One way pionered by~\cite{goga2013large} is to use a binary classifier predicting if two profiles match; profile similarity in this case would be the confidence of positive class. Learning to rank methods~\cite{trotman2005learning} is another way to cast profile matching as a supervised problem. The supervised models provide a convenient framework where name similarity features, used in our method, can be mixed with attribute-, network-, and  image-based features.

\section*{Acknowledgements}

This research was conducted as part of a project funded by Digital Society Laboratory LLC. We thank Prof. Chris Biemann and three anonymous reviewers for their thorough comments that significantly improved quality of this paper.

\bibliographystyle{splncs}
\bibliography{sigproc}

\end{document}